\documentclass[epj]{svjour}
\usepackage{amsfonts}
\usepackage{amssymb}

\usepackage[pdftex]{graphicx}
\usepackage{color}
\usepackage[pdftex,
           breaklinks=true,
           colorlinks=true,
           ]{hyperref}

\newcommand{\Zn}{ZnCu$_3$(OH)$_6$Cl$_2$}

\begin{document}

\title{Magnetic susceptibility and specific heat of the spin-$\frac{1}{2}$
Heisenberg model  on  the kagome  lattice and  experimental data on {\Zn}.}
\authorrunning{G. Misguich and P. Sindzingre}
\titlerunning{Magnetic susceptibility and specific heat of the spin-$\frac{1}{2}$ Heisenberg model ...}

\abstract{
We compute the magnetic     susceptibility and specific heat  of   the
spin-$\frac{1}{2}$  Heisenberg  model    on  the  kagome  lattice with
high-temperature  expansions and exact  diagonalizations.  We  compare
the  results with the  experimental data on   {\Zn} obtained by Helton
{\it  et al.} [Phys.   Rev. Lett. {\bf  98},  107204 (2007)].  Down to
$k_BT/J\simeq0.2$, our      calculations   reproduce accurately    the
experimental    susceptibility,    with an    exchange     interaction
$J\simeq190$~K  and  a  contribution of  3.7\%   of weakly interacting
impurity  spins.    The comparison  between   our  calculations of the
specific heat and  the  experiments indicate that  the low-temperature
entropy   (below $\sim$20~K) is  smaller  in {\Zn} than  in the kagome
Heisenberg  model, a likely signature   of  other interactions in  the
system.
\PACS{
      {75.50.Ee}{Antiferromagnetics}   \and
      {75.10.Jm}{Quantized spin models} \and
      {75.40.Cx}{Static properties}
     } 
}

\author{Gr\'egoire Misguich\inst{1} \and  Philippe Sindzingre\inst{2}}

\institute{Service de Physique Th\'eorique, CEA Saclay, 91191 Gif-sur-Yvette Cedex, France
\and
Laboratoire de Physique Th\'eorique de la Mati\`ere
condens\'ee, Univ. P. et M. Curie, 75252 Paris Cedex, France}
\mail{gregoire.misguich@cea.fr}
\maketitle

\section{Introduction}
After many years of theoretical
investigations, the nature of the ground-state of the spin-$\frac{1}{2}$ Heisenberg model
on the kagome lattice is still not known.
Although all numerical studies have concluded to the absence of long-range
magnetic (N\'eel) 
order \cite{elser89,ze90,ce92,sh92,le93,nm95,lecheminant97,web98},
many basic questions such as
the existence of spontaneously broken symmetries, or the existence of a finite
gap to magnetic excitations remain open.
In fact, many different states of matter have been proposed
for the kagome Heisenberg antiferromagnet: 
$\mathbb{Z}_2$ gapped topological liquids \cite{sachdev92,wv06},
valence-bond crystals \cite{mz91,hastings00,ns03,ba04},
critical spin liquids with gapless spinons \cite{ran06,hastings00}.

Recently, a promising spin-$\frac{1}{2}$ antiferromagnetic
insulator with an ideal kagome geometry, {\Zn}, has
been synthesized and studied for its magnetic 
properties \cite{helton06,mendels06,ofer06,imai07,devries07}.
Because these studies did not detect any kind ordering (nor spin freezing) down
to 50mK, it could represent one of the first and most remarkable realization 
of a 2D quantum spin liquid \cite{anderson73,ml05}.

To extract some information about the low-energy physics
of the kagome Heisenberg model from the experiments on {\Zn},
it is important to first analyze in a quantitative way the possible
role of magnetic 
defects (and other ``perturbations''to this model) in this compound.
In this paper we compare the experimental data for
the magnetic susceptibility  $\chi$ and  specific heat $c_v$
obtained by Helton {\it et al.}\cite{helton06}
with calculations  for the spin-$\frac{1}{2}$ Heisenberg model 
on the kagome lattice based on exact diagonalization (ED) data 
(partial spectrum of a 36-site cluster and
full spectrum for 24-site and 18-sites clusters) 
and high-temperature series expansion \cite{ey94,mb05}.
Down to temperature $k_BT/J=0.2$, the experimental susceptibility 
$\chi_{\rm exp}(T)$ can be very well fitted by 
that of  the kagome lattice  Heisenberg model with $J\simeq190$~K
plus  a contribution of about 4\%  of  impurity spins with weak mutual 
interactions
(modeled by a ferromagnetic Curie-Weiss temperature of $\simeq-6$~K),
likely mostly due to antisite disorder 
(Cu substituted to Zn on sites between the kagome planes) \cite{devries07,SHlee97}.
The low  temperature specific  heat is dominated by impurities 
(and other perturbations)
below 2~K and by phonons above 15~K. In the intermediate range,
the calculated specific heat appears to be larger than in the experiment.
We comment on this feature at the end of the paper.

\section{Uniform static susceptibility}
\label{sec:chi}

The spin-$\frac{1}{2}$ Heisenberg  model reads:
\begin{equation}
    \mathcal{H}=J\sum_{\langle i,j\rangle} \vec S_i\cdot\vec S_j -g\mu_BH\sum_i S_i^z
    \label{eq:H}
\end{equation}
where the sum runs over pairs of nearest neighbor sites on the kagome lattice
and $H$  is an  external   magnetic  field.
To fix the notations, we define the (zero-field)
uniform susceptibility per  site $\chi(T)$ as
$
    \chi(T)=\frac{g\mu_B}{N}\left. \frac{\partial  \sum_i \langle S_i^z \rangle
}{\partial H}\right|_{H=0}
$ where $N$ is the total number of spins.

\begin{figure}
\includegraphics[width=8cm]{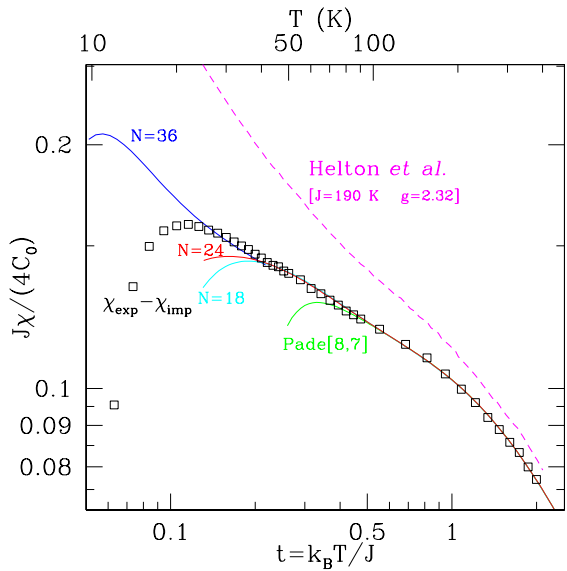}\vspace*{-2cm}
\caption{
(color online)
Magnetic susceptibility per spin as a function of
temperature. Dashed (magenta) curve~: experimental data (Helton {\it et al.}),
multiplied  by $J/(4C_0)$ as a function of $k_BT/J$  with
$J\simeq190$~K  and $C_0=0.504$~K~cm$^3$/(mol of Cu).  
Black squares: 
$\chi_{\rm s}=\frac{J}{4C_0}(\chi_{\rm exp}(T)-\chi_{\rm imp}(T))$, 
obtained from the experimental data $\chi_{\rm exp}$  
by   subtracting the contribution $\chi_{\rm imp}$ of a 
concentration $x=0.03655$ of impurity spins with a Curie-Weiss temperature 
$\theta_{\rm imp}=-6.1$~K.  Red (resp. cyan)
curve : Exact $\chi_{\rm th}$  for a $N=24$ (resp. 18)  site kagome
cluster   with  periodic boundary  conditions.  Blue curve:
Results for $N=36$ spins obtained with the (approximate) method of
reference \cite{sindzingre00}.
Green curve: Pad\'e approximant from the high-temperature expansion at order $t^{-15}$.
The Pad\'e approximant is not converged below $t\simeq 0.4$
whereas the finite-size curves are practically converged to the thermodynamic
limit down to $t\simeq0.2$. Below $t=0.2J$, the later curves are only indicative.
 } \label{fig:chi}
\end{figure}

The  high-temperature expansion of $\chi$ has been computed
up   to    order $\mathcal{O}\left(T^{-15}\right)$ (included) 
by    Elstner  and Young~\cite{ey94}:
\begin{eqnarray}
         \chi(T)&=&\frac{4C_0}{J}\chi_{\rm th}(t=k_BT/J) \nonumber \\
    \chi_{\rm th}(t)&=&
    (1/4)t^{-1}-(1/4) t^{-2}+\cdots
\end{eqnarray}
where $C_0=0.25(g\mu_B)^2$ is the Curie constant and $t$ the reduced temperature.
The truncated series  $\chi_{n}(t)=\sum_{i=0}^{n}c_i t^{-i}$ at  order
$n=14$ and  $n=15$ agree with  a relative error smaller than $10^{-2}$
for $t>1$. Down to this temperature, they already provide good
approximations to $\chi_{\rm th}(t)$. 
The convergence
of high-$T$ series  can be improved  using Pad\'e  approximants (PA).
In the present  case
the  PA provide a reliable estimate of $\chi_{\rm th}(t)$ at least down  to
$t\sim0.5$.\footnote{
Comparing the PA and the exact
curves for 18 and 24-site clusters shows that the PA is in fact correct
down to $T\sim 0.4J$, see Fig.~\ref{fig:chi}.
}
One representative PA (numerator
of degree 8 and denominator of degree 7) is  displayed
Fig.~\ref{fig:chi}. 

On a small enough system,  it is possible to obtain by ED the {\it full}
spectrum ($2^N$ energy levels for $N$ spins).
We  have done so for two 18-site and 24-site kagome clusters 
(with periodic  boundary conditions). 
Then thermodynamic quantities  can be computed exactly 
as a function of $T$.  
For bigger systems, where one can still compute some eigenstates by ED, 
one may use the approximate method  described  in
reference \cite{sindzingre00}  to compute thermodynamic quantities.
footnote{
In this   method, one constructs the density of states in each symmetry sector 
from  the exact low and high energy  states obtained from ED
and approximating in between the unknown part of the spectrum with a
smooth density of states.
This smooth part is constructed so that the first moments of the
density of states ($Tr[H^n]$),
in each symmetry sector of the finite cluster,
are exact up to $n=5$.
This Ansatz guaranties that the thermodynamics
becomes exact at low T (when thermal excitations
only involve the eigenstates computed exactly) as well as at high T
(when a re-summed high-temperature expansion up to $T^{-5}$ is valid).
}

The  results for the susceptibility  $\chi(t)$ are shown in  Fig.~\ref{fig:chi}.
Above $t=0.2$, the relative difference between the (exact) $N=18$ and 
$N=24$ curves is smaller than $0.5\%$. 
We therefore make the (rather safe) assumption that our finite-size results
are good approximations to the thermodynamic limit down to 
$t_{\rm min}\simeq0.2$.
This represents a small gain over the
coupled-cluster expansion of reference \cite{rs07}, 
which is  valid above $t\sim0.3$.

The $\chi(t)$  obtained  with the approximate method for $N=36$ sites 
also agrees (with a relative error smaller than 2\%) with  the $N=24$ results 
down to $t\simeq0.2$.
Slightly below, the 36-site susceptibility  is still increasing 
and might be a better approximation to the infinite-size limit than the
24-site curve. 
Still, it is not  possible decide at which $t$ 
finite-size (and/or errors due to the approximation in the
density of states) will become too important.
Safely, we only use the theoretical results (noted $\chi_{\rm th}$)
above $t\simeq0.2$ to fit the experimental data $\chi_{\rm exp}$
for the susceptibility.

We fit $\chi_{\rm exp}$ 
to a sum of contributions from the kagome spins $\chi_{\rm s}$
and the impurities $\chi_{\rm imp}$ in the following way:
\begin{eqnarray}
    \chi_{\rm exp}(T)=\frac{4C_0}{J}\chi_{\rm s}(k_BT/J)+\chi_{\rm imp}(T) \\
{\rm with}\;\;
	\chi_{\rm imp}(T)=\frac{xC_0}{T+\theta_{\rm imp}}
	\label{eq:chi}
\end{eqnarray}
where $J$ is the (unknown)  magnetic exchange in  between the spins
in the kagome  planes, $x$ an impurity   concentration,
$C_0=0.25(g\mu_B)^2/k_B$
the Curie constant and $\theta_{\rm imp}$ the Curie-Weiss temperature of the  system of impurities.
Eq.~\ref{eq:chi} is the leading term in a high-temperature expansion for the
system of impurities,  
which provides a simplified picture of their interactions. 
To be applicable, $T$ should therefore be large compared to $\theta_{\rm imp}$. 
We assume that the system of impurities does not perturb the the kagome spins.

We optimized  numerically the parameters so that $\chi_{\rm s}$ 
fits the theoretical results $\chi_{\rm th}$ for $t\ge 0.2$.  
As can be  seen on  Fig.~\ref{fig:chi}, an  almost perfect agreement
can be obtained between 
$\chi_{\rm s}=\frac{J}{4C_0}(\chi_{\rm exp}(T)-\chi_{\rm imp}(T))$
(squares) and the theoretical estimates 
for the  kagome susceptibility $\chi_{\rm th}$ 
with the following   parameters: $C_0=0.504$~K~cm$^3$/(mol of Cu)
(equivalent to a gyromagnetic factor $g=2.32$),  $J=190.4$~K,
$x=0.03655$ and $\theta_{\rm imp}=-6.1$~K. 
We note that the value of $J$ is in rough agreement with the values reported 
in reference \cite{helton06} (17meV$\simeq$200~K) and  \cite{rs07} (170~K). 
We also check a posteriori that the lowest temperature of the fit
($t=0.2\simeq38$~K) is much bigger than $|\theta_{\rm imp}|$, so that a 
Curie-Weiss approximation for the impurities is justified. 
6~K is  also approximatively the transition temperature
reported in Zn$_{x}$Cu$_{4-x}$(OH)$_6$Cl$_3$ for $x<0.6$ (replacing some Zn by Cu between the kagome planes) \cite{mendels06},
so this energy scale may correspond to some couplings  for spins located between the planes, where magnetic impurities could sit.
We eventually notice that down to
$t=0.15$ (that is below the lowest temperature used for
the fit),   $\chi_{\rm exp}-\chi_{\rm imp}$
continues to increase and to follow the $N=36$ curve.
This suggests that the maximum of the kagome susceptibility could indeed 
be below $t=0.15$.

Rigol and Singh \cite{rs07} analyzed the same experimental data 
with another high-temperature method (coupled cluster expansion) and
obtained a somewhat different conclusion.
They argued that Dzyaloshinskii-Moriya (DM) interactions
provide a better description of the low-temperature increase of the susceptibility than impurities.
Although we agree that DM interactions are certainly present in {\Zn} and that they should affect the  physics of the system (at least at low-temperatures),
it is also clear that a few percent of impurities must be present too and should have a visible effect on the susceptibility, even at rather high temperatures.
According to reference \cite{rs07}, {\it free} impurities cannot explain 
the sharp increase of $\chi_{\rm exp}$ below 60~K. In our analysis, this issue 
is solved by allowing for a small {\it ferromagnetic} Curie-Weiss temperature 
$\theta_{\rm imp}\simeq -6$~K for the impurities.\footnote{
The Curie-Weiss temperature is an {\it average} of the different exchange 
constants.  However, due to the  complexity of the interactions between 
impurities (disorder), the ferromagnetic sign of $\theta_{\rm imp}$ does not 
necessarily imply that they behave ferromagnetically at {\it low} temperatures.
}
One can indeed see that, once $\chi_{\rm imp}$ has been subtracted,
the experimental results show a saturation of $\chi$ around $20$~K,
in rough agreement with
the measurements of reference \cite{imai07}.
The location of the maximum we obtain is however quite sensitive to the value $x$ of the  impurity concentration.

\section{Specific heat}
The experimental data for the specific heat $c_v(T)$
are only available at very low temperature where the 
size effects on the ED results are large.  
We therefore also applied the  high-temperature 
entropy method \cite{bm01}.
It combines  three  pieces of information  about  the system:
1) The high temperature series expansion of  $c_v(T)$,
up to $T^{-17}$ \cite{ey94,mb05}.
2) The  ground-state energy  per site $e_0$ of the Hamiltonian. 
Here  we use the following estimate 
$e_0=2\langle 0|\vec  S_i\cdot\vec S_j|0\rangle=-0.44$ \cite{mb05}.
3) The exponent $\alpha$  describing the low-temperature behavior of  the
specific heat: $c_v(t\to 0)\sim t^\alpha$.
The method  then provides a set  of $c_v(t)$ curves 
(for  different Pad\'e  approximants)
which all satisfy exactly the following properties: i) $c_v(T\to 0)\sim
T^\alpha$, ii) $c_v(T\to \infty)\sim$ the series expansion,
iii) $\int_0^\infty c_v(T)dT=-Ne_0$, and  iv)
$\int_0^\infty c_v(T)/TdT=Nk_B\ln(2)$.
When the value of $e_0$ and  $\alpha$ are both correct, 
one  usually gets a  large  number  of  very similar curves
but if either is {\em incorrect}, only a few and scattered curves 
will be obtained (more details in Refs.~\cite{bm01,mb05}).  
In the case of   the  kagome antiferromagnet, the value of $\alpha$  
is not known  and the entropy method   gives a  reasonable  convergence  
for  $\alpha=1$    and $\alpha=2$ \cite{mb05}.
Motivated  by the experimental  observation \cite{helton06} of a
low-temperature  $c_v$  with an  exponent  smaller than one, we
also include here a calculation with $\alpha=0.5$.

\begin{figure}
\includegraphics[width=8cm]{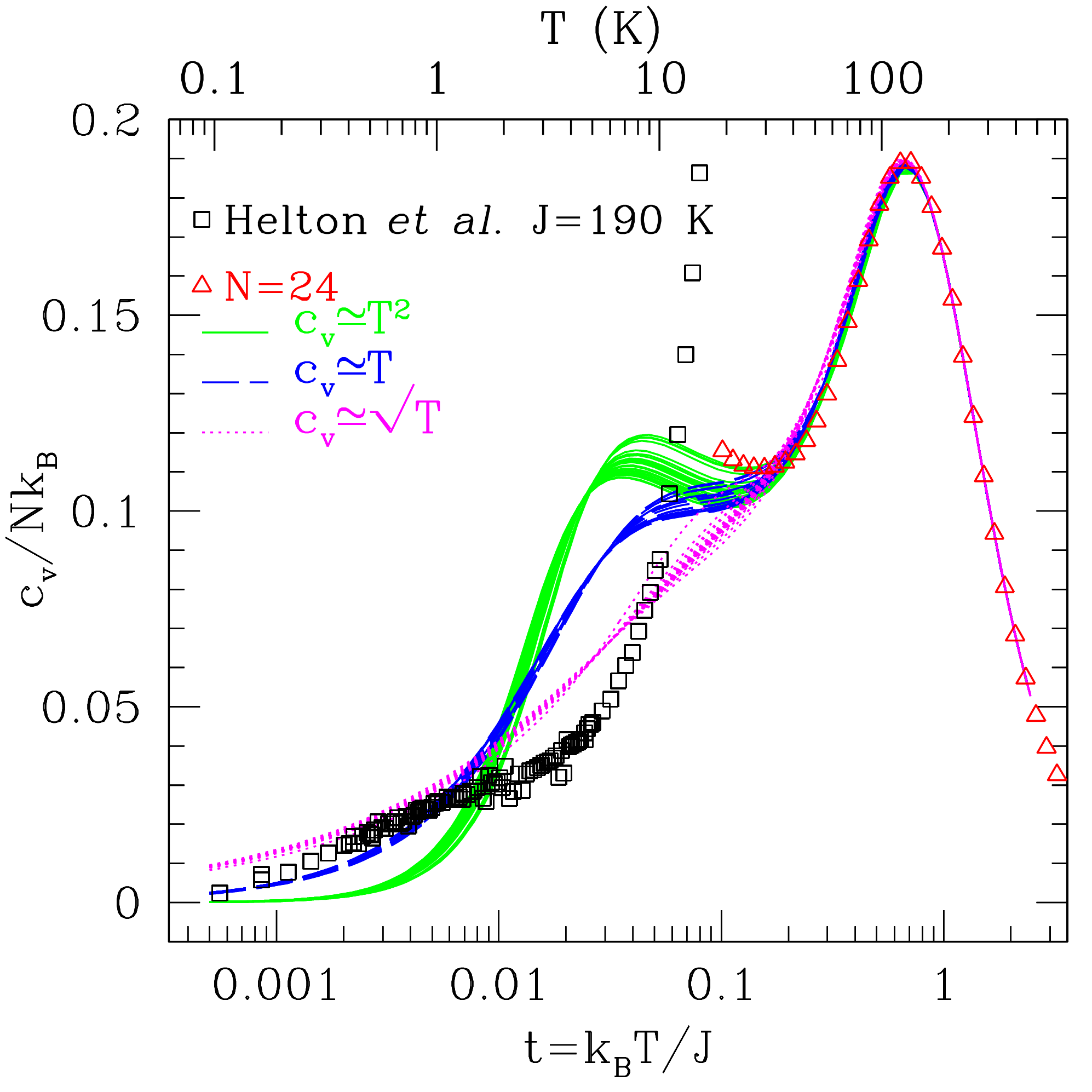}
\caption{
(color online)
Specific heat $c_v$ vs temperature $T$.
Black
squares:
experimental data from Helton {\it et al.} \cite{helton06}, assuming
$J\simeq190$~K. Red triangles: exact  specific heat of a  24-site kagome cluster.
Green (full), blue (dashed) and magenta (dotted) curves: 
$c_v$ calculated by the entropy method for three  
different values  of the  low-temperature exponent
($\alpha=2$, 1 and 0.5).
All  valid PA from order 14 to 17 with numerator and denominator of degrees 
$\geq3$ are shown.
For a given $\alpha$ and at each temperature, 
their dispersion provides a rough estimate of the error bars.}
\label{fig:cv}
\end{figure}

The  results for some valid\footnote{By the entropy method,
the  entropy $s(e)$ is obtained  as the power $\alpha/(\alpha+1)$ of a rational fraction
(PA) of the  energy per site $e$.   The  specific  heat curve is then
obtained     parametrically      through     $T(e)=1/s'(e)$        and
$c_v(e)=-s'(e)^2/s''(e)$.  Only the PA which satisfy $s(e)>0$, $s'(e)>0$ and
$s''(e)<0$  in  the range   $]e_0,0[$  are  physically
``valid''.} PA from orders 14 to 17 are displayed
in Fig.~\ref{fig:cv}
together with the exact result obtained by ED 
of the full spectrum of a  24-site  kagome cluster.
For $t>0.3$, these results  are practically exact.

In the two scenarios $\alpha=1$ and 2 there is a significant dispersion of the
results for  $t<0.3$ from one PA to another \cite{mb05}.\footnote{This is due to the finite order in the high-temperature expansion. 
Still, for a given  value of $\alpha$,  we  believe  that  this  method  
gives a qualitatively correct  picture  for $c_v(T)$, 
even at low T.
}
In  both case, $c_v$  show a low-t peak or shoulder 
as found from  ED of finite-size systems.
The choice $\alpha=0.5$ leads to a smoother $c_v(t)$ and
improves significantly the convergence.
It is however not clear if this improved convergence
for small values of $\alpha$ ($0.5\lesssim\alpha\lesssim1$) 
indicates that $\alpha$ is actually smaller than one or
an artifact of the present entropy method, which might be
``slower'' to stabilize a low-t peak 
(as with $\alpha=2$, see  Fig.~\ref{fig:cv}), than a smooth curve
(as with $\alpha=0.5$). 
In any case, this is clearly related to the unusually large 
low-temperature entropy of the kagome system.

\begin{figure}
\includegraphics[width=8cm]{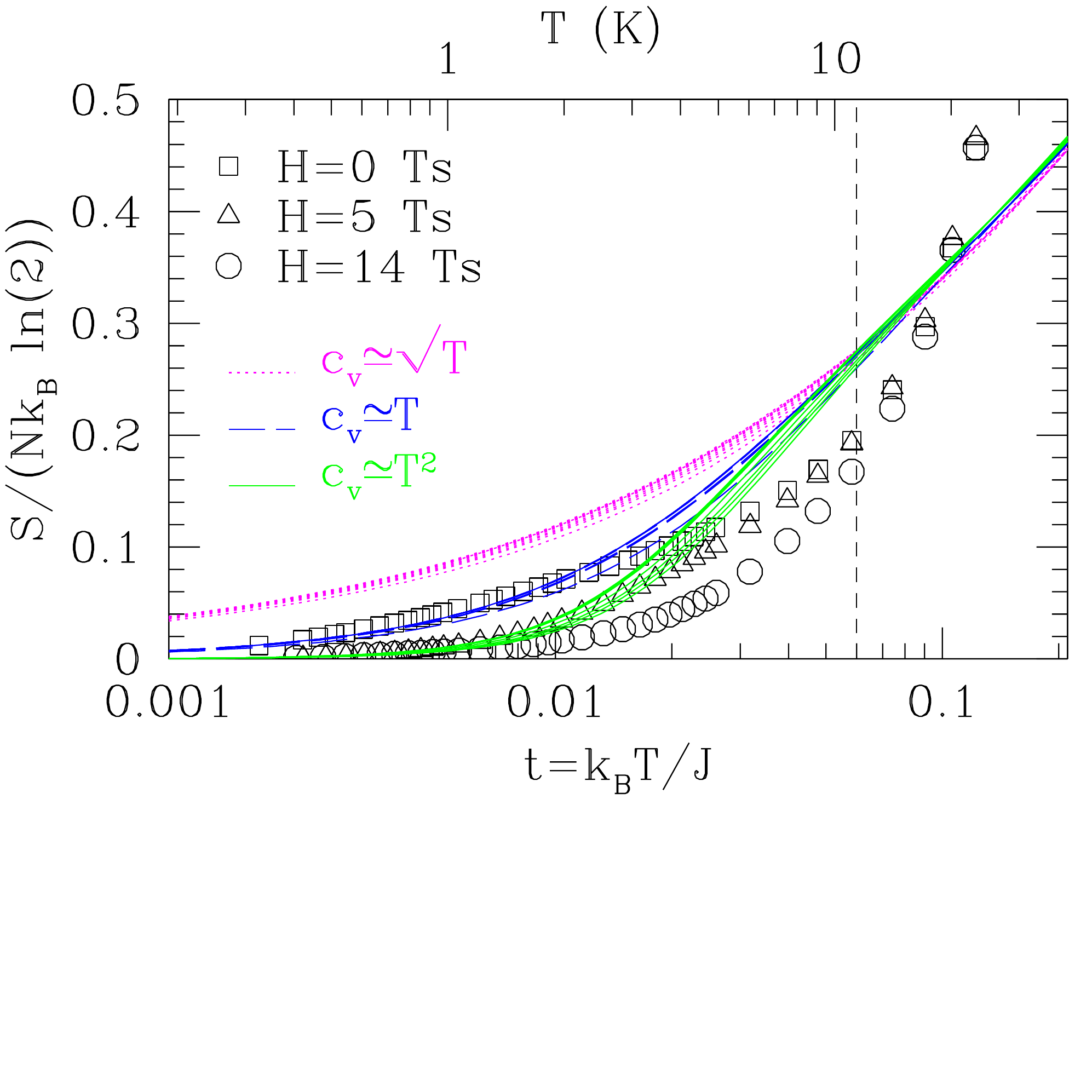}\vspace*{-2.3cm}
\caption{
(color online)
Entropy $S$ vs temperature.
Square, triangles and circles:
experimental data from Helton {\it et al.} \cite{helton06}
with  magnetic field B= 0, 5 and 14 Teslas, plotted as a function of $k_BT/J$ with $J=190$~K. 
Green (full), blue (dashed) and magenta (dotted) curves: entropy calculated by the
entropy method for $\alpha=2$, $\alpha=1$ and $\alpha=0.5$ (same PA as in Fig.~\ref{fig:cv}).}
\label{fig:s}
\end{figure}

Fig.~\ref{fig:cv} also displays the experimental results
(black squares) obtained by Helton {\it et al.} \cite{helton06}.
The only  parameter here is the exchange
constant,  taken to be $J\simeq190$~K  from  the fits of the susceptibility data.
Above 15~K, the phonon contribution is dominant and we  have to focus
on the lower temperatures to analyze the magnetic contribution.
Below 15~K the order of magnitude agrees with our calculation but
there is no quantitative agreement.
Several ``perturbations'' such as weakly interacting magnetic impurities 
or magnetic anisotropies should  indeed contribute to the specific heat 
at such low temperatures.

Results for the integrated entropy  $S(T)=\int_0^T c_v(x)/xdx$ 
are plotted  in Fig.~\ref{fig:s},
where one sees  that the choice of the low-$T$ exponent 
$\alpha$ of $c_v$ has practically no influence on the 
theoretical $S(t=k_BT/J)$ above $t\simeq0.06$ and that,
around this temperature, the experimental value 
is significantly {\it lower} than in our calculations.
Of course, subtracting the contribution of the phonons 
(hard to estimate quantitatively)
and from the impurities would make the discrepancy even larger.
Concerning the impurities, one sees in Fig.~\ref{fig:s}
that an applied magnetic field of 5 and 14 Teslas is enough to
significantly reduce $S(t)$ for $t\lesssim0.06$.
Such fields are low in comparison to $J$ but of the order of the
estimated coupling between the impurities.
The difference between the curves at 0 and 5 (or 14) Teslas may thus provide a rough
estimate of the contribution of the impurities.
Note that $S(t)$ at 5
Teslas become closer to the $S(t)$ computed
by the entropy method with $\alpha=2$.
The experimental value $\alpha<1$ could be due to the impurities and
the actual value of $\alpha$ for the kagome spins could be larger than 1.
In any case,
at $t=0.06$($\simeq 11$~K), the experimental entropy ($\simeq 0.2\ln{2}$) is thus at 
{\it least} 7\% of $\ln{2}$ below the theoretical estimates ($0.27\ln{2}$).
This seems 
a rather robust indication
that some additional interactions  play some role in this energy range, 
by freezing some degrees of freedom of the spins in the kagome planes 
and pushing the corresponding entropy  to higher energies. 
We may mention in particular DM
interactions \cite{rs07},
non-magnetic impurities in the kagome planes (``dilution'') \cite{devries07}
and interactions  between impurities and 
the kagome spins.
We conclude that 15$\sim$20~K is a minimal temperature
for a kagome lattice Heisenberg model description of
{\Zn} to be valid.

\section*{Acknowledgments}
We are grateful to C.~Lhuillier for 
many discussions
and comments about the manuscript.
We also thank P.~Mendels
and F.~Bert for useful discussions.
GM also thanks Y.~S.~Lee and J.~Helton for discussions and for providing their data
as well as P.A.~Lee, Y.~Ran, T.~Senthil, X.-G. Wen for discussions on related topics.

Note added : After the first submission of this manuscript, two preprints \cite{devries07,SHlee97}
(neutron scattering) 
confirmed the importance  of magnetic impurities (from  6\% to 10\%)
in {\Zn}. The  smaller value found here
could be due to our simplified model to fit the magnetic
susceptibility.

\end{document}